\documentclass[aps,preprint]{revtex4}
\usepackage{graphicx}
\usepackage{bm}

\begin{document}
\date{\today}
\title{S-matrix theory of single-channel ballistic transport through coupled
quantum dots}
\author{ I. Rotter$^1$ and A.F. Sadreev$^{2,3}$}
\affiliation {$^1$ Max-Planck-Institute f\"ur Physik komplexer
Systeme, D-01187 Dresden, Germany } \affiliation{$^2$  Kirensky
Institute of Physics, 660036, Krasnoyarsk, Russia}
\affiliation{$^3$ Department of Physics and Measurement
Technology, Link\"{o}ping  University,  S-581 83 Link\"{o}ping,
Sweden}

\begin{abstract}
We consider  single-channel transmission through a double quantum
dot system that consists of two single dots  coupled by a wire of
finite length $L$. In order to explain the numerically obtained
results for a realistic double dot system we explore a simple
model. It consists, as the realistic system, of two dots connected
by a wire of length $L$. However, each of the two single dots is
characterized by a few energy levels only, and the wire is assumed
to have only one level whose energy depends on the length $L$. The
transmission is described by using $S$ matrix theory. The model
explains in particular the splitting of the resonant transmission
peaks and the origin of the transmission zeros. The latter are
independent of the length of the wire.
When the transmission zeros of the single dots are of first order
and both single dots are identical, those of
the double dot are of second order.  First-order transmission
zeros cause phase jumps of the transmission amplitude by $\pi$,
while there are no phase jumps related to second-order
transmission zeros. In this latter case, a phase jump appears due
to a resonance state whose decay width vanishes when crossing the
energy of the transmission  zero.

\end{abstract}

\maketitle

\section{Introduction}

Advances in nanotechnology have made it possible to fabricate
quantum dots (QDs) and to study their transport properties. Single
quantum dots  can be considered as artificial atoms if the energy
levels can be resolved. Two or more QDs can be coupled to form an
artificial molecule, in which the electrons are shared by
different sites. Recently, much interest is devoted to the study
of the properties of these coupled QDs or artificial molecules
\cite{bird}. One of the most important challenges is to understand
the basic properties of coupled QDs since they display the
simplest structures of quantum-computing devices that can be
controlled by means of external parameters. An example is the
interdot coupling that can be tuned by external parameters far out
of the regimes known  for natural molecules. The ground-state
interatomic distance is dictated, in natural molecules,  by the
nature of bonding \cite{rontani} while such a restriction does not
exist in artificial molecules. Most of the work has focused on the
situation where the coupled QDs are in the tunneling regime so
that the physics is dominated by Coulomb blockade effects
\cite{blick,livermore}.  It was observed a pronounced
anticrossing  resonance scenario  (called also avoided level
crossing scenario) instead of a simple crossing one.
Recently Rushforth {\it et al.} \cite{rushforth} studied  the
movement of electrons between the dots. Application of a voltage
onto the gate between the dots allows to vary the length as well as
the width of the wire connecting the dots. The study  revealed evidence
that electrons move between the dots via excited states of either the
single dots or the double dot molecule.

In the present paper, we consider the ballistic transport through
a double QD in the regime where Coulomb
blockade effects can be neglected and where the transmission is
resonant. The two single QDs are connected by
a wire of finite length. This gives the possibility to vary the
length or the width of the wire and to study the transmission
through the double QD as a function of the wire's size and
energy. When the length $L$ of the wire is much larger than its
width then the wire has at least one eigenmode with the energy
$\epsilon \propto L^{-2}$. This mode appears additionally to the
eigenstates $\varepsilon \propto R^{-2}$ of the two single QDs
(where $R$ is the radius of the dot) which may be equal to one
another or different from one another. Such a double QD system allows
therefore to investigate, among others, the quantum mechanical
problem of the coupling of two identical quantum systems that is
provided by a third quantum mechanical subsystem with an own
energy spectrum. This system is the analogue of a
molecule with hydrogen bonds.

Some years ago, the phase of the transmission amplitude  has been
measured in a double-slit interference experiment \cite{heilblum}.
The results showed  phase  jumps by $\pi$ between resonances which
raised intensive theoretical  work for an explanation
\cite{zeros,lee,kim,mois,guevara,silva}. Most of these
calculations associate the sharp phase drops
with the occurrence of transmission zeros and relate them
to the interference zeros of
Fano resonances.  In \cite{mois}, it was shown however  that the
existence of a transmission zero is, indeed, a necessary condition
for the phase jump but not a sufficient one. The  sharp phase
change bases, according to \cite{mois},  on the destructive
interference between neighboring resonance states and thus differs from
the mechanism based on the Fano interference picture. Destructive
interferences between neighbored resonances are considered also in
\cite{silva,kim}.

The Fano resonance phenomenon characterizes the interference
between a single resonance with a relatively smooth background
\cite{fano}. The interference processes in the regime of
overlapping resonances are, however, much more complicated than
those in the regime of isolated resonances. This has been
demonstrated, e.g., in an experimental study of the conductance
through a quantum dot in an Aharonov-Bohm interferometer
\cite{kob} and in a theoretical study \cite{marost}. These results
are another hint to the conclusion, drawn in \cite{mois}, that the
sharp phase changes observed in \cite{heilblum} are the result of
processes being different from the simple Fano interference
picture.

Here, we study the transmission properties of the double QD system when
one lead is attached to the first single QD,  another one
to the second single QD, and both single QDs are connected by a wire of
finite length $L$. Also in this system, transmission zeros appear.
The special situation of a double QD is such that
the transmission zeros of the whole system are determined by the
zeros in the transmission through the single QDs.
Since phase jumps in the transmission amplitudes are related to the
transmission zeros, the mechanism of their appearance
in a double QD system is expected to be
different from that based on the simple picture of Fano resonances.

We will show in the present paper that transmission zeros of first
order of a single 2d QD cause  phase jumps by $\pi$ in the amplitude
of the transmission through this single QD. It depends on the
spectral properties of both single QDs and on the manner they are
connected to the wire and to the leads, whether or not the
transmission zeros of the single QDs cause phase jumps of the
transmission amplitudes  of the double QD system.  When the single
QDs remain true 2d-systems in the double QD (as shown in Fig.
\ref{fig4}) and have different energy spectra, each transmission
zero of each single QD causes a transmission zero of the same type
in the double QD system.  When the spectra of both single QDs are
however equal (and their connection to the wire and the leads is
the same as above), the corresponding transmission zeros of the
double QD system are of second order. They give rise to two phase
jumps, each by $\pi$, that compensate each other. When a resonance state
crosses this transmission zero at a certain length of the wire, its decay
width vanishes  and a phase jump appears now due to the extremely
narrow resonance. The transmission zeros do not depend
on the length of the wire when there is only one channel for the propagation
of the mode in the wire and in the leads.

In Sect. II, we present some numerical results for the
transmission through a double QD that consists of two single QDs
connected by a wire. The main features of the transmission are
represented as a function of the length $L$ of the wire inside the
double QD: the transmission zeros  are independent of $L$ while
many transmission peaks show some periodicity as a function of
$L$. In the following sections, we study the transmission through
a double QD in detail by using  a simple model with only a few
levels in both single QDs and one level in the wire. The
description is based on the $S$ matrix theory for the transmission
through quantum dots \cite{sadreev}. In Sect. III, the formalism
is derived and some typical numerical results are given and
discussed. The spectral properties of the double QD are characterized
by the (complex) eigenvalues of the effective Hamiltonian that describes
the double QD when opened by attaching the two leads to it. This
effective Hamiltonian is non-Hermitian, and its eigenvalues
provide both the positions in energy and the decay widths of the states.
The appearance of transmission zeros is determined by the spectral
properties of the two single QDs.

The relation between transmission zeros and phase jumps is
discussed in Section IV. Most results are obtained for the case that the
two single QDs keep their 2d structure when connected to the leads and to
the wire. The results show very clearly  that the
mechanism  differs from that based on the simple Fano interference
picture: the spectral properties of the states of the double QD system
depend strongly on the length of the wire, while the transmission
zeros do not depend at all on the length. The transmission zeros are
of first order
when the energy spectra of the two single QDs are different from
one another, and of second order when the spectra of the two single QDs
are identical.
With the last section, we conclude the present study by mentioning
some results for more complicated double QD systems. In particular
we  consider the case that the widths of the leads and of the wire
are different from one another. When the wire's width exceeds the
lead's one, a few channels can propagate through the wire and the
transmission zeros depend on the wire's length.


\section{Single channel transmission through double dots}

\begin{figure}[ht]
\includegraphics[width=.8\textwidth]{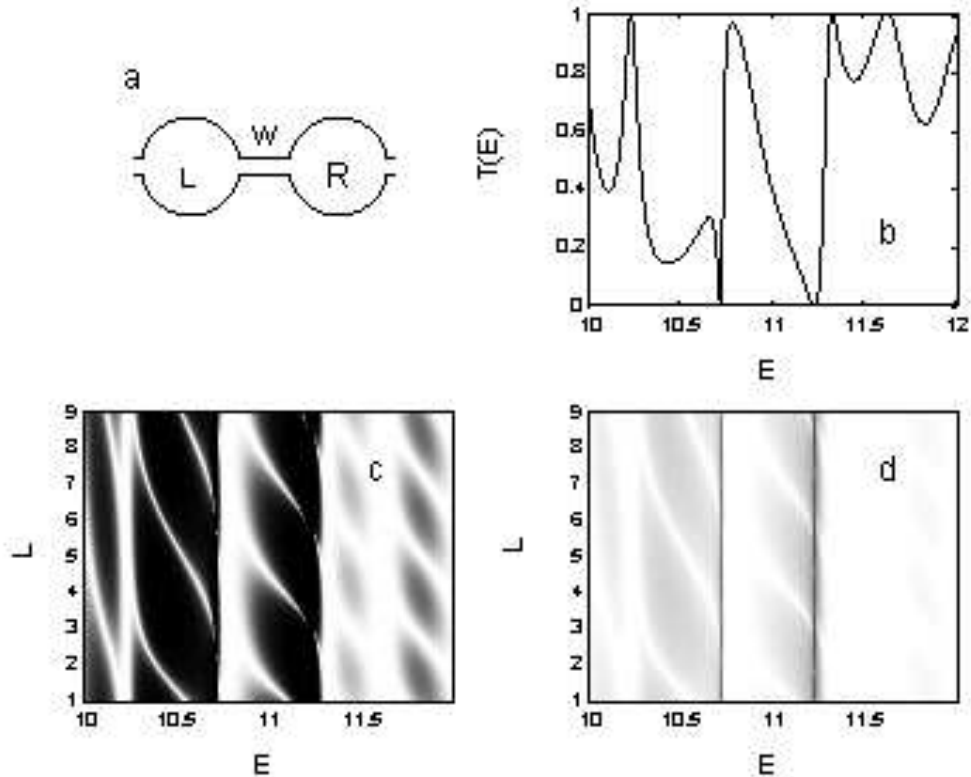}
\caption{ (a) The shape of the double QD that consists of two
identical circular dots  and a wire connecting the two dots.
Numerical size of the double QD: radius $R=40$ and width of the
wire and leads  $d=8$. (b) The probability $T_1(E)$ for the
transmission through one of the single circular dots versus
energy. (c) The probability $T(E,L)$ for transmission through the
double QD shown in (a) versus energy and length of the wire. (d)
The same as (c) but $ln(T(E,L))$ in order to show clearly the
zeros for the transmission through the double QD.} \label{fig1}
\end{figure}

Let us consider first a double QD consisting of two identical
single QDs, for example, two circular dots, that are connected by
a wire of the length $L$. For simplicity we choose the width of
the wire $d$ equal to the width of the leads, see Fig. \ref{fig1}
(a). The theory of transmission through such a system is given by
Klimenko and Onipko \cite{klimenko}. If the dots are identical,
the probability of the transmission for the single-channel case is
\begin{equation}\label{T2}
  T=\frac{T_1^2}{T_1^2+4(1-T_1)\sin^2(\phi+kL)}
\end{equation}
where $T_1=|t_1|^2, \quad t_1$ is the complex amplitude of the
transmission through the single QD, $\phi=arg(t_1)$, and $k$ is
the wave number related to the energy of the single-channel
transmission as
\begin{equation}\label{kE}
  E=k^2+\pi^2.
\end{equation}
All values are dimensionless via the characteristic energy
$\frac{\hbar^2}{2m^{*}d^2}$ and the width $d$ of the wire and the
leads. In the expression (\ref{T2}), evanescent modes are ignored
whose wave vectors  are imaginary. That means, (\ref{T2}) can be
used for $kL\gg 1$. The  results shown in Fig. \ref{fig1} are
obtained from numerical calculations with $k>3$. Here, formula
(\ref{T2}) is applicable already for $L>1$. Similar results for a
system of two single QDs with different shape were obtained
numerically by Pichugin \cite{pichugin}.

The transmission probability demonstrates a few interesting
features that are shown in Fig. \ref{fig1} (b) - (d). The first is
some periodical dependence of the transmission peaks on the wire
length and on the energy. This dependence can be seen immediately
in formula (\ref{T2}), and we will not discuss it further. The
second feature are the transmission zeros. They do not at all
depend on the wire length, see Fig. \ref{fig1} (d) where the
transmission probability $T$ is shown in logarithmic scale. Also
this feature follows directly from (\ref{T2}):  the zeros of the
transmission probability $T_1$ through the single QD lead directly
to zeros of the total transmission probability $T$. The third
feature concerns the peaks of the transmission probability.
Fig. \ref{fig1} demonstrates crossings and anticrossings of
resonant peaks as observed in experiments \cite{blick,rushforth}.
Some of them are independent of the length $L$, as the
transmission zeros, while other ones are dependent on $L$. The
latter ones cross  the $L$ independent peaks as well as the zeros,
but the behaviour in the vicinity of the crossing with a zero is
different from that in the vicinity of the crossing with a
maximum.

In the following, we will consider these features in detail.
For this purpose, we use the periodicity (first feature) of the
transmission picture that allows us to restrict the investigation to the
transmission properties of a simple model with only a few states.
The study is based on the S-matrix theory.

\section{S-matrix description of the transmission through double dots}

\subsection{Closed system consisting of two single dots connected by a wire}

The Hamiltonian of the double QD shown in Fig. \ref{fig1} (a)
consists of three parts: two parts describe the two single QDs and
a third one is related to the wire. The Hamiltonian of the two single
QDs is formulated in standard way,
\begin{equation}\label{Hdots}
  H_d=H_L+H_R=\sum_{n_L=1}^{N_L}\varepsilon_{n_L}|n_L\rangle \langle n_L| +
  \sum_{n_R=1}^{N_R}\varepsilon_{n_R}|n_R\rangle \langle n_R|\; ,
\end{equation}
where the indices $L, R$ stand, respectively, for the left and
right single QD with the energies $\varepsilon_{n_L},
\varepsilon_{n_R}$ and the  Hilbert dimensions $N_L, N_R$.
The wire is the third independent quantum mechanical subsystem
described by the Hamiltonian
\begin{equation}\label{Hwire}
H_w=\sum_{n_w=1}^{N_w}\epsilon_{n_w}(L)|n_w\rangle \langle n_w| \;
.
\end{equation}
We assume that the eigenenergies $\epsilon_{n_w}(L)$ of the wire
depend only on its length $L$. This offers the possibility for the
energies of the wire to cross the eigenenergies of the single QDs.
We assume further that the wire is coupled to, respectively, the
left and right single QD via the matrices $U_L, ~U_R$ of rank
$N_L\times N_w,\; N_R\times d_w$. Then the total Hamiltonian has
the following matrix form
\begin{equation}\label{HB}
  H_B=\left(\begin{array}{ccc} H_L    & U_L & 0 \cr
         U_L^{+} & H_w & U_R^{+} \cr
            0  &   U_R &  H_R   \end{array}\right).
\end{equation}
The Hamiltonian (\ref{HB}) differs from those  used in the
literature \cite{kisilev,kim,guevara} for the description of a
double QD of similar shape by taking explicitly into account the
third part (\ref{Hwire}) for the wire.

For the simplest case  $N_L = N_w = N_R = 1$ and equal single QDs,
the total Hamiltonian takes the following form
\begin{equation}\label{HB3}
  H_B=\left(\begin{array}{ccc} \varepsilon_1    & u & 0 \cr
         u & \epsilon(L) & u \cr
            0  &  u &  \varepsilon_1  \end{array}\right).
\end{equation}
The eigenvalues of this Hamiltonian are
\begin{eqnarray}\label{ED3}
  E_{1,3}=\frac{\varepsilon_1+\epsilon(L)}{2} \mp \eta,
\quad  E_2 = \varepsilon_1,
  \\
  \eta^2=\Delta\varepsilon^2 + 2u^2, \; \;\Delta\varepsilon=
  \frac{\varepsilon_1 - \epsilon(L)}{2}
\end{eqnarray}
and the eigenstates read
\begin{equation}\label{psiB3}
  |1\rangle =\frac{1}{\sqrt{2\eta(\eta+\Delta\varepsilon)}}\left(\begin{array}{c} -u\cr
           \eta+\Delta\varepsilon \cr
            -u \end{array}\right),
            |2\rangle =\frac{1}{\sqrt{2}}\left(\begin{array}{c} 1 \cr
           0 \cr    -1 \end{array}\right),
  |3\rangle =\frac{1}{\sqrt{2\eta(\eta-\Delta\varepsilon)}}\left(\begin{array}{c} u\cr
           \eta-\Delta\varepsilon \cr
            u \end{array}\right).
      \end{equation}
It is remarkable that one of the eigenenergies of the total system
coincides with the energy $\varepsilon_1$ of the single QD. This
fact remains also in the more general cases with higher
dimensions.

Let us consider the case of two identical single QDs with the
number $N$ of states  that are coupled via the wire. Then it
follows $U_L = U_R$, if the wire is a straight one. The
determinant which defines the eigenvalues of the total Hamiltonian
is of rank $2N+N_w$,
\begin{equation}\label{det}
  \left|\begin{array}{llllllll}
E_1-E & 0      &\cdots & U_{11} & U_{12}&\cdots & 0     &   0 \cr
  0   & E_2-E  &\cdots & U_{21} & U_{22}&\cdots & 0     &   0 \cr
\vdots& \vdots &\vdots &\vdots  &\vdots &\vdots &\vdots &\vdots\cr
U_{11}&U_{21}  &\cdots &\epsilon_1(L)-E&0 &\cdots &U_{12}
&U_{11}\cr U_{12}&U_{22}  &\cdots &0&\epsilon_2(L)-E &\cdots
&U_{12} &U_{11}\cr \vdots& \vdots &\vdots &\vdots &\vdots &\vdots
&\vdots &\vdots\cr 0     & 0      &\cdots & U_{21}& U_{22} &\cdots
& E_2-E & 0  \cr 0     & 0      &\cdots & U_{11}& U_{12} &\cdots &
0 &  E_1-E \cr
\end{array}\right|=0.
\end{equation}
When $E=E_n,\, n=1,\ldots, d,$ two lines in (\ref{det})  coincide.
Therefore, among the $2N+N_w$ eigenenergies of the Hamiltonian
(\ref{HB}), there are $N$ eigenvalues which are equal to the
eigenenergies of the single dots. Numerical examples for the
eigenvalues of $H_B$ with $N=1$ and 2 are given by the solid lines
in Fig. \ref{fig3}.

\subsection{S-matrix for the transmission through
double dots: single QDs with one state.}

\begin{figure}[ht]
\includegraphics[width=.8\textwidth]{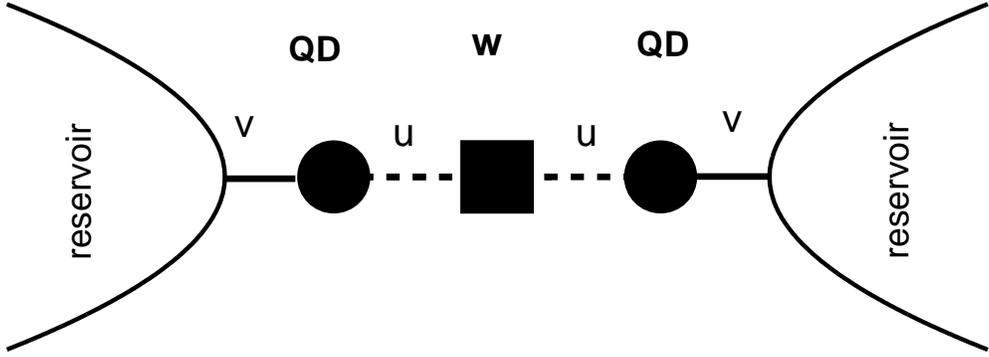}
\caption{Two single state  QDs are connected to the wire w
with the
coupling constants $u$ and to the reservoirs with the coupling
constants $v$.} \label{fig2}
\end{figure}

The knowledge of the eigenstates of the closed quantum system
allows us to formulate the S-matrix and the effective Hamiltonian
in the manner described in \cite{sadreev}. Let us consider first the
most simple case of the Hamiltonian (\ref{HB3}). For the transmission
through this system we imply two leads coupled to the reservoirs
with the strength $v$, as shown in Fig. \ref{fig2}.

The coupling matrix has the following general form
\cite{dittes,sadreev}
\begin{equation}\label{Vmatrix}
  V=\sum_m \sum_{C=L,R}\int dE \; V_m(E,C) |E,C\rangle \langle m| +H.C.
\end{equation}
where $|m\rangle $ are the eigenstates of the closed system given in
the present case by (\ref{psiB3}), and $C$ enumerates the reservoirs with
the states $|E,C\rangle $  normalized by
$$\langle E,C|E',C'\rangle =\delta(E-E')\delta_{C,C'}.$$
Obviously,
\begin{equation}\label{VME}
V_m(E,C)=\langle E,C|V|m\rangle =
\sum_j\int_C dx \, \psi_C(x) \psi_m(j)\, \langle x|V|j\rangle
\end{equation}
where $j$ runs over the double dot system, $x$ spans over the C-th
reservoir, $\psi_C(x)$ are the eigenfunctions of the reservoirs,
and $\psi_m(j)$ are the eigenfunctions of the closed  double QD
system. In our model we choose the couplings, that are shown in Fig.
\ref{fig2} by solid lines, to be fixed at some points: at $j=1,
~3$ of the left and the right single QD and at some points $x_C$
which belong to the reservoirs. Moreover, we describe for
simplicity the reservoirs as  semi infinite one-dimensional wires
in tight-binding approach \cite{sadreev}. As in \cite{sadreev}, we
take the connection points of the coupling to the reservoirs at
the edges of the one-dimensional leads. Then the matrix elements
(\ref{VME}) take the following form
\begin{eqnarray}
\label{Vm} V_m(E,L)=v\psi_{E,L}(x_L)\psi_m(j=1)=v\sqrt{\frac{\sin
k}{2\pi}}\psi_m(1),\nonumber\\
V_m(E,R)=v\psi_{E,L}(x_R)\psi_m(j=3)=v\sqrt{\frac{\sin k}{2\pi
}}\psi_m(3),
\end{eqnarray}
where $k$ is the wave vector related to energy by $E=-2\cos k +2$.
For continual case the last equality is simply $E\approx k^2$. The
effective Hamiltonian can be written as
\cite{dittes,sadreev}
\begin{equation}
\label{Heffgen} H_{\rm eff}=H_B+\sum_{C=L,R}
V_{BC}\frac{1}{E^{+}-H_C}V_{CB},
\end{equation}
where $H_C$ is the Hamiltonian of the reservoir $C$ and $E^{+}=E+i0$.
Substituting (\ref{Vm}) into (\ref{Heffgen}) we obtain for the matrix
elements of the effective Hamiltonian
\begin{eqnarray}\label{Heff}
 \langle  m|H_{\rm eff}|n\rangle & =& E_m\delta_{mn}+
\sum_{C=L,R}\frac{1}{2\pi}\int_{0}^4 dE'
\frac{V_m(E',C)V_n(E',C)}{E+i0-E'}\nonumber\\
&=& E_m\delta_{mn}-v^2(\psi_m(1)\psi_n(1)+\psi_m(3)\psi_n(3))e^{ik},
\end{eqnarray}
where the states $\psi_n(j)$ are given in (\ref{psiB3}) and the
indices $j=1, 3$ mean, respectively, the left single QD $(j=1)$
and the right single QD $(j=3)$. Substituting (\ref{psiB3}) into
(\ref{Heff}) we obtain
\begin{equation}\label{Heff3}
H_{\rm eff}=\left(\begin{array}{ccc}
E_1-\frac{v^2u^2e^{ik}}{\eta(\eta+\Delta\varepsilon)}& 0 &
\frac{v^2ue^{ik}}{\sqrt{2}\eta} \cr 0& \varepsilon_1-v^2e^{ik} & 0
\cr  \frac{v^2ue^{ik}}{\sqrt{2}\eta} & 0 &
E_3-\frac{v^2u^2e^{ik}}{\eta(\eta-\Delta\varepsilon)}\cr
             \end{array}\right) \; .
\end{equation}
Next we  calculate the (complex) eigenvalues of
the effective Hamiltonian (\ref{Heff3}) that are related to
the poles of the $S$ matrix. The result is
\begin{eqnarray}\label{poles3}
z_2 & = & \varepsilon_1-v^2e^{ik},\nonumber\\
z_{1,3} & = & \frac{\varepsilon_1+\epsilon(L)-v^2e^{ik}}{2} \mp
  \sqrt{\left(\frac{\epsilon(L)-\varepsilon_1+v^2e^{ik}}{2}\right)^2+2u^2}.
\end{eqnarray}
For $u=0$, the eigenvalues are $z_{1,3}=\varepsilon_1-v^2e^{ik},
z_2=\epsilon(L)$. The eigenvalue $z_2=\epsilon(L)$ means that the wire
has no other connection to the reservoirs as that via the single
QDs.

In order to calculate the S-matrix  we need the eigenstates of the
effective Hamiltonian \cite{sadreev}
\begin{equation}\label{Heffeigstates }
  H_{\rm eff}|\lambda)=z_{\lambda}|\lambda),
\end{equation}
where $(\lambda|=|\lambda)^{c}, \lambda = 1, 2 , 3$, $c$ means
transposition, and $(\lambda|\lambda')=\delta_{\lambda,\lambda'}$
is the biorthogonality relation for the eigenfunctions of the non-Hermitian
Hamiltonian  $H_{\rm eff}$ \cite{rotter}.
Then if follows from (\ref{Heff3}) (for $u\neq 0$):
\begin{eqnarray}\label{psiHeff3}
|1)=\left(\begin{array}{c}
  a \cr   0\cr   b\end{array}\right), \; \;
|2)=\left(\begin{array}{c} 0 \cr
           1 \cr    0 \end{array}\right), \; \;
|3)=\left(\begin{array}{c}b \cr 0\cr -a
             \end{array}\right)
\end{eqnarray}
with
\begin{eqnarray}
             a=-\frac{f}{\sqrt{2\xi(\xi+\omega)}},\;\;
b=\sqrt{\frac{\xi+\omega}{2\xi}}
      \end{eqnarray}
and
\begin{equation}\label{Delxi}
  f=\frac{v^2ue^{ik}}{\sqrt{2}\eta}, \;
  \omega=-\eta+\frac{\Delta\varepsilon v^2e^{ik}}{2\eta}, \;
  \xi^2=\omega^2+f^2.
\end{equation}

The knowledge of the eigenstates (\ref{psiHeff3}) of the effective
Hamiltonian  allows us to write the amplitude for the transmission
through the double QD in simple form \cite{sadreev},
\begin{equation}
\label{trHeff3} t=-2\pi
i\sum_{\lambda}\frac{\langle L|V|\lambda)(\lambda|V|R\rangle }{E-z_{\lambda}}.
\end{equation}
The transmission probability is $T=|t|^2$.
Substituting (\ref{Vm}), (\ref{psiHeff3}) and correspondingly
(\ref{psiB3}) into the  matrix elements $\langle L|V|\lambda)$ and
$(\lambda|V|R\rangle $ we obtain
\begin{eqnarray}\label{Vlam}
\langle L|V|2)&=&\sum_m\langle E,L|V|m\rangle \langle
m|2)=\frac{v}{2}\sqrt{\frac{\sin k}{\pi}},\nonumber\\
(2|V|R\rangle &=&\sum_m(2|m\rangle \langle m|V|E,L\rangle
=-\frac{v}{2}\sqrt{\frac{\sin k}{\pi}},\nonumber\\ \langle
L|V|1)=(1|V|R\rangle & =& v\sqrt{\frac{\sin
k}{2\pi}}(\psi_1(1)a+\psi_3(1)b), \nonumber\\ \langle
L|V|3)=(3|V|R\rangle &=&v\sqrt{\frac{\sin
k}{2\pi}}(\psi_1(1)b-\psi_3(1)a),
\end{eqnarray}
where the eigenstates $\psi_m(j)$ are given in (\ref{psiB3}).
Substituting finally (\ref{Vlam}) into (\ref{trHeff3}) we obtain
the transmission amplitude $t$.

\begin{figure}[ht]
\includegraphics[width=.8\textwidth]{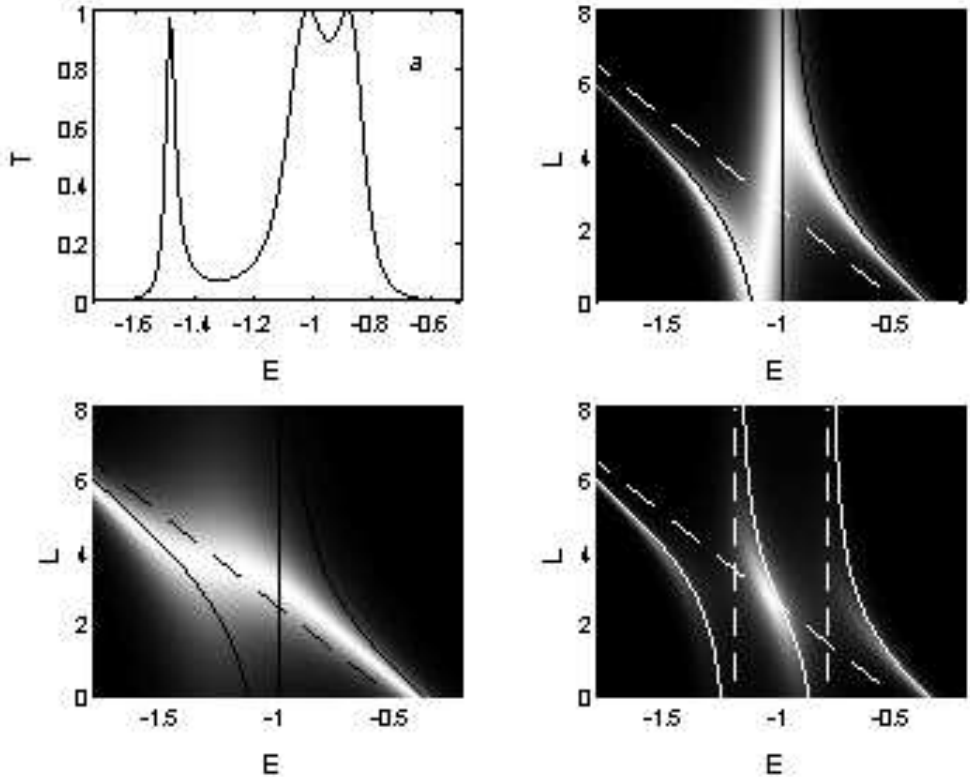}
\caption{The transmission probability $T$ through the double QD
shown in Fig. \ref{fig2}. (a) $\varepsilon_1=-1, ~v=0.3, ~u=0.2$
and $L=4$. (b) The same as (a) but the length $L$ is not fixed.
(c) The same as (b) but $\varepsilon_1=-1, ~v=0.6, ~u=0.2$. (d)
The same as (b) but $\varepsilon_1^L=-1.2, ~\varepsilon_1^R=-0.8,
~v=0.3, ~u=0.2$. The eigenenergies $E_k(L), \; k=1,2,3,$ of $H_B$
are shown by full lines while the  energies  $\varepsilon_{1}$ and
$\epsilon(L) = -1/2-L/5$ are given by dashed lines.} \label{fig3}
\end{figure}

The typical behavior of the transmission
probability $T=|t(E,L)|^2$ versus energy $E$ and length $L$ is
shown in Fig. \ref{fig3}.
Here the energies $\varepsilon_k$ of the two single QDs and the
energy $\epsilon (L)$ of the wire are shown by dashed lines while
the eigenenergies (\ref{ED3}) of the double QD system are shown by solid
lines. Since the eigenenergy $E_2$ of the double QD system coincides with
the energy $\varepsilon_1$ of the single QDs, the last one is not
shown in Fig. \ref{fig3} (b). The eigenenergy of the wire
is assumed to depend on the length $L$ according to
$\epsilon(L)=-1/2-L/5$. This linear dependence is not decisive
for the following discussion, see the Concluding Remarks.

We underline that the results presented in   Fig. \ref{fig3}
follow from a simple model which describes the double QD system by
two single-state dots connected by a wire. The wire  is
characterized by the only energy $\epsilon(L)$. When the coupling
of the double QD system to the reservoir is relatively weak
[meaning that the ratio $v/u$ is small as in Fig. \ref{fig3} (a,
b, d)], the transmission probability follows the eigenenergies of
the closed double QD, and we have resonant transmission. We can
now compare Fig. \ref{fig3} (b) with Fig. \ref{fig1}, and  we see
that the simple model, basic of Fig. \ref{fig3} (a), is able to
explain the peaks of the transmission probability. That means, the
main features
of the transmission picture in Fig. \ref{fig1}, are related to the
crossing of the levels of the single QD with the energy level of
the wire.  As shown in Fig. \ref{fig3} (b), (c) and (d)
different scenarios of the resonant transmission peaks can be
realized from the crossing to the anticrossing behavior in
dependence on the parameters, basically on  the coupling strength
between the system and the attached leads and on the coupling strength
between the dots and the wire.

We can further learn from Fig. \ref{fig3} that the transmission
through a system consisting of two dots that are connected by a
wire,  is characterized by the ratio $v/u$. Here, $v$ is the
coupling strength of each single QD to the corresponding lead and
$u$ is the internal bonding of the two single QDs. Thus, $v/u$ is
the ratio between external and internal coupling of the states of
the double QD via, respectively, the reservoir and the wire. In
Fig. \ref{fig3} (c), the coupling $v$ between the double QD and
the reservoirs exceeds essentially  the coupling $u$ of the two
single QDs to the wire. In this case,  the transmission is mainly
given by the resonant transmission through the wire, and the two
single QDs become parts of the reservoirs as it is directly seen
from (\ref{poles3}).

Moreover, Fig. \ref{fig3} (d) shows  the transmission through a
double QD system  consisting of two different single QD  connected by a
wire for the small ratio $v/u$. Also in this case, the
transmission probability completely follows  the eigenenergies of
the double QD.

All the results obtained for a double QD with one-site single QDs
do not show any transmission zeros in energy. In order to
demonstrate the absence of zeros, we plotted in Fig. \ref{fig3}
(a)  the energy dependence of the transmission probability for an
arbitrary but fixed length $L$ of the wire. The  model underlying
the results of Fig. \ref{fig3} can be considered as a
one-dimensional chain of three sites.
It is in complete agreement with the consideration by Lee \cite{lee}
that odd and even resonance levels alternate in energy in
realistic 1d systems so that  zeros in the transmission
probability can not appear. The results of Fig. \ref{fig3}
correlate also with the consideration of a simple two site system
\cite{guevara,sadreev}. It has been shown for these systems that
an architecture of the couplings between system and reservoirs,
which violates the true one dimensionality of the closed system,
gives rise to a zero in the transmission probability at a certain
energy.

\subsection{S-matrix for the transmission through
double dots: single QDs with many states.}

\begin{figure}[ht]
\includegraphics[width=.8\textwidth]{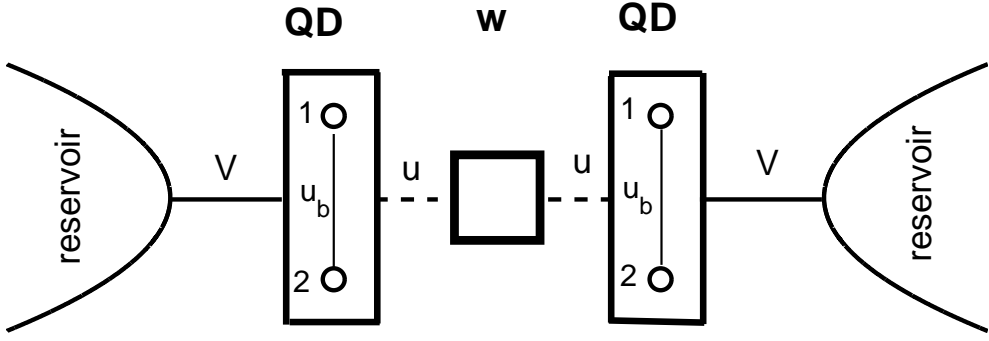}
\caption{The double dot system is connected to the reservoirs by
the coupling constants $v$. The single dots are coupled to the
wire by the coupling constants $u$.} \label{fig4}
\end{figure}

Here, we consider the transmission through a double QD when each
single QD of the system is presented by two states as shown in
Fig. \ref{fig4}.
The Hamiltonian (\ref{HB}) of the closed double QD consisting of
the two single QD and the wire is
\begin{equation}\label{HB5}
H_B=\left(\begin{array}{ccccc}\varepsilon_1 & 0 & u & 0 & 0   \cr
                               0 & \varepsilon_2 &  u & 0 & 0 \cr
                               u &  u & \epsilon(L) & u & u \cr
0 & 0 &  u & \varepsilon_2 & 0 \cr
            0  &  0 & u & 0 & \varepsilon_1  \end{array}\right).
\end{equation}
For simplicity we assume that all the coupling constants between
the wire and the single QD are the same and are given by the
constant value $u$.

 The Hamiltonian (\ref{HB5}) is written in the energy
representation (\ref{Hdots}), (\ref{Hwire}). In order to specify
the connection between the reservoirs and the single QDs, we have
however to know   the eigenstates of (\ref{HB5}) also in the site
representation. The Hamiltonian of the single QD in the site
representation is
\begin{equation}\label{Hd2}
H_b=\left(\begin{array}{cc}\varepsilon_0 & u_b \cr
                               u_b & \varepsilon_0\cr
                                 \end{array}\right) \; .
\end{equation}
The hopping matrix elements $u_b$ are shown  in Fig.
\ref{fig4} by thin solid lines. The eigenfunctions and
eigenvalues are the following
\begin{eqnarray}\label{Hd2eigs}
\langle j|\varepsilon_1\rangle & = &
\frac{1}{\sqrt{2}}\left(\begin{array}{c}1 \cr
                               1\cr \end{array}\right), \; \;
\langle j|\varepsilon_2\rangle =\frac{1}{\sqrt{2}}\left(\begin{array}{c}1 \cr
                               -1\cr
                               \end{array}\right),\nonumber\\
\varepsilon_{1,2} & = & \varepsilon_0 \mp u_b.
\end{eqnarray}
We introduce the projection operators
\begin{equation}\label{P5}
  P_L = \sum_{b_L}|\varepsilon_{b_L}\rangle \langle \varepsilon_{b_L}| \; ,
  \quad P_w= |1_w\rangle \langle 1_w| \; , \quad
  P_R=\sum_{b_R}|\varepsilon_{b_R}\rangle \langle \varepsilon_{b_R}|
\end{equation}
where $b_L = 1, 2$, ~$ b_R = 1, 2$, and  $|1_w\rangle $ is the
one-dimensional eigenstate of the wire. Let $E_m$ and $|m\rangle $
with $m=1, ..., 5$ denote  the five eigenenergies and eigenstates
of (\ref{HB5}), $H_B|m\rangle =E_m|m\rangle $. The  elements of
the left coupling matrix are
\begin{eqnarray}\label{VL5}
\langle L,E|V|m\rangle & = &\langle L,E|V P_L
|m\rangle =\sum_{b_L}\langle L,E|V|\varepsilon_{b_L}
\rangle \langle \varepsilon_{b_L}|m\rangle
\nonumber\\
& =& \sum_{j_L=1,2}\sum_{b_L}\langle L,E|V|j_L\rangle
\langle j_L|\varepsilon_{b_L}\rangle \langle \varepsilon_{b_L}|m\rangle .
\end{eqnarray}
Similar expressions can be derived for the right coupling matrix.
Here we used the assumption that the left reservoir is connected
only to the left single QD and the right reservoir only to the
right single QD. As previously,  the reservoirs are assumed to be
semi infinite one-dimensional wires. Next we have to specify which
sites of the left (right) single QD are connected to the left
(right) reservoir. There are two possibilities.

(i) Assume the left reservoir is connected only to the first site
$j_L = 1$ of the left single QD. Then (\ref{VL5}) becomes with
account of (\ref{Hd2eigs})
\begin{equation}\label{VLj1}
\langle L,E|V|m\rangle =v\sqrt{\frac{\sin k}{2\pi}}
\sum_{b_L}\langle \varepsilon_{b_L}|m\rangle .
\end{equation}
A corresponding expression can be written down for the right
coupling matrix if the right reservoir is connected to the first
site of the right single QD.

(ii) We can assume that the reservoirs are connected to both sites
of the single QDs with the same coupling constant $v$. Then the
elements of the coupling matrices (\ref{VLj1}) are the following
\begin{equation}\label{VLj2}
\langle L,E|V|m\rangle = v\sqrt{\frac{\sin k}{2\pi}}\langle
\varepsilon_1|m\rangle
\end{equation}
provided that the energy level $\varepsilon_1$ is the lowest in energy,
see (\ref{Hd2eigs}).

The most important difference between the previous
$d=1$ case  and the present $d=2$ one for the single QD is that the
system is now no longer  necessarily
one dimensional. Therefore, zeros in the transmission
probability may appear.

As it was shown above, Eq. (\ref{det}), two eigenvalues of the
Hamiltonian $H_B$ coincide with the energies $\varepsilon_1$ and
$\varepsilon_2$ of the single QD. The other three eigenvalues of
(\ref{HB5}) can be found by solving of a cubic equation. Also the
finding of the eigenstates of (\ref{HB5}) is a formidable task. In
what follows, we consider therefore the transmission through a
system with two states of each single QD numerically.

\begin{figure}[ht]
\includegraphics[width=.8\textwidth,height=0.4\textheight]{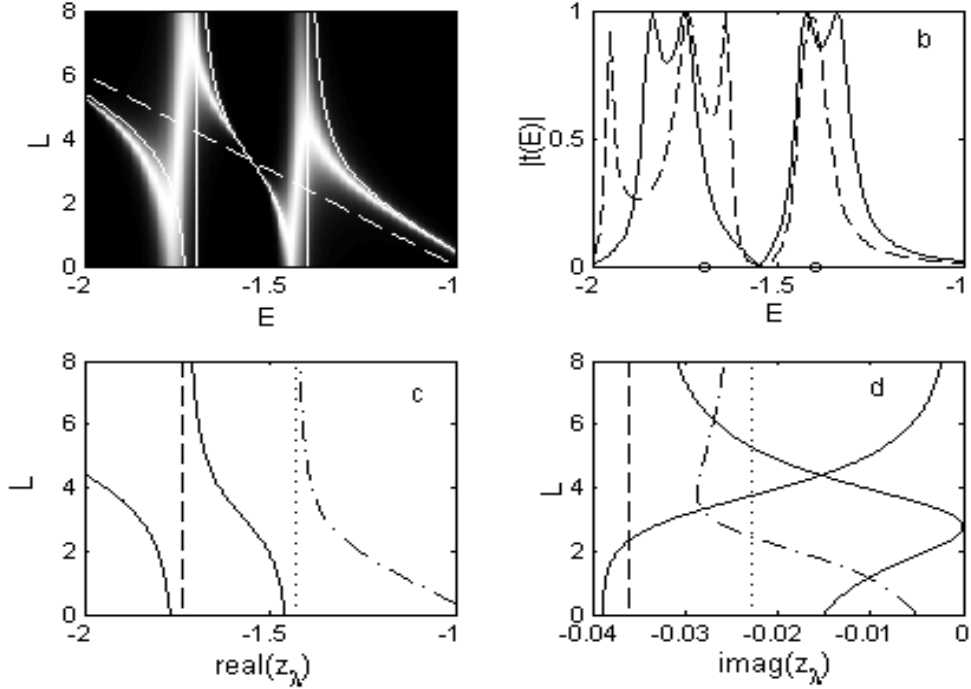}
\caption{ (a) The transmission through a double QD  with two
identical single QDs that are connected by a wire according to
Fig. \ref{fig4}. The eigenvalues of $H_B$ are shown by full lines.
$\varepsilon_1 = -1.7,~ \varepsilon_2 = -1.4$ and $~\epsilon(L) =
-1 - L/5$ (dashed line), ~$v = 0.3, ~u = 0.1$. (b) The modules of
the transmission amplitude  $|t(E,L)|$ for the same double QD as
in (a) for fixed lengths $L = 2.75 $ (solid line) and $L = 4$
(dashed line). The energies of the two single QDs are shown by
circles. The real part (c) and imaginary part (d)  of the 5
eigenvalues $z_k$ of the effective Hamiltonian as a function of
$L$ for $E=-1.5$. Thin solid line:  $z_1$, dashed line: $z_2$,
thick solid line: $z_3$, dotted line:  $z_4$, and dash-dotted
line: $z_5$. At $L = 2.75 $  the imaginary part of the third
eigenvalue is equal to zero at all energies $E$. } \label{fig5}
\end{figure}
In  Fig. \ref{fig5}, the transmission probability versus energy
$E$ and length $L$ of a double QD  is shown for the case that both
sites of the single QD are connected to the reservoir with the
coupling matrix elements (\ref{VLj2}). The figure shows a  zero in
the transmission probability, indeed, see Fig. \ref{fig5} (b).
According to Figs. \ref{fig5} (c) and (d), the positions and decay
widths of the eigenstates 2 and 4  of the effective Hamiltonian
are independent of the length $L$ of the wire while those of the
other states depend on $L$. The state 3, lying  in the middle of
the spectrum, crosses the transmission zero at  $L = 2.75$. Here,
the decay width of this state approaches zero for all energies $E$.
The transmission zeros
will be discussed in detail in the next section. Here, we remark
only that resonance states with vanishing decay width are considered
also by other authors. In \cite{guevara}, they are called ghost
Fano resonances that appear in a double quantum dot molecule
attached to leads. In atomic physics, the phenomena related to
resonance states with vanishing decay width are known as population
trapping \cite{marost2}. They  result from the interplay of the
direct coupling of the states and their coupling via the continuum
under the influence of, e.g., a strong laser field.

\section{Zeros in single-channel transmission }

\begin{figure}[ht]
\includegraphics[width=.8\textwidth,height=0.4\textheight]{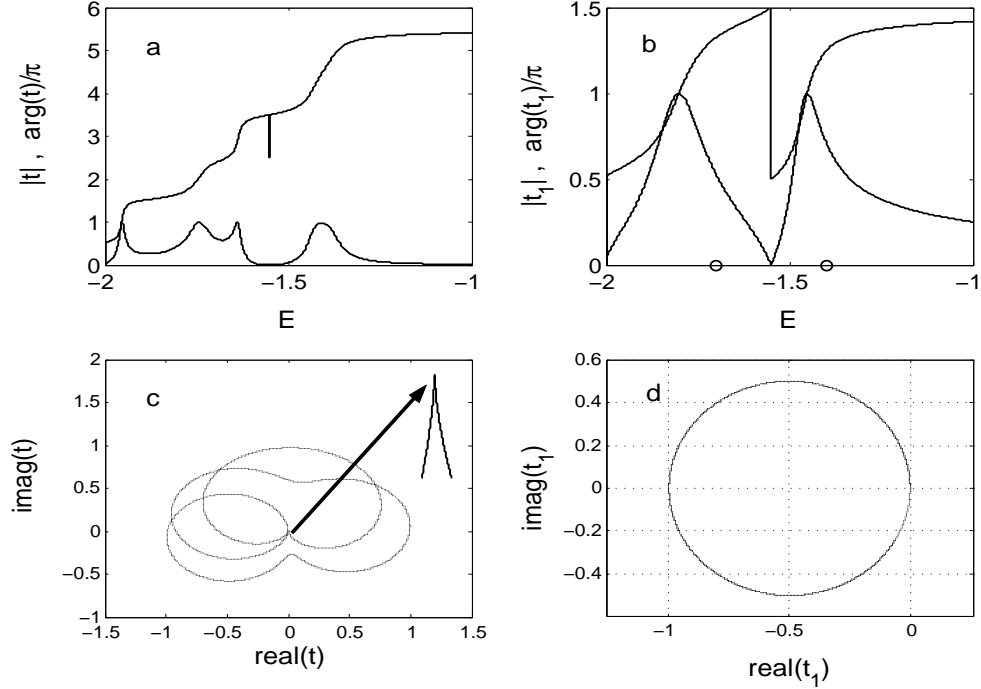}
\caption{(a) The modules  $|t(E)|$ (lower curve)
and the phase $arg(t)/\pi$ (upper curve) of the transmission amplitude
for a double QD with two identical single QDs.
The parameters are the same as in Fig.
\ref{fig5}, and $L=4$ as in Fig. \ref{fig5} (b),  dashed line. (b)
The modules   $|t_1(E)|$ (lower
curve) and the phase $arg(t_1)/\pi$ (upper curve) of the transmission
amplitude  for  one of the single QDs
that is part of the double QD considered in (a). The energy positions of
the single QD levels are shown by circles at the abscissa. (c)
Evolution of imaginary and real parts of the transmission
amplitude $t$ of the double QD with energy. At the upper right
corner, a zoomed fragment of the evolution is shown which
demonstrates that the evolution has cusp-like behavior in the
vicinity of the transmission zero. (d) The same as (c) but for the
single QD. Here, the evolution is of standard type. } \label{fig6}
\end{figure}

In Fig. \ref{fig6} (a),   the transmission through a double QD system
with two identical single QDs is shown, while   Fig. \ref{fig6}
(b), shows the transmission through one of these single QDs (the lower
curves correspond to the modules of the transmission amplitude
and the upper curves to
their phases). In the double QD system, the two single QDs are
connected to the leads and to the wire as shown in Fig. \ref{fig4}.
Comparing the two figures, we see that the
transmission zero of the double QD coincides with that of the
single QD. This result is in agreement with formula (\ref{T2}) for
the single-channel transmission through identical dots. However
there is a remarkable difference between the zeros in both cases
as will be explained in the following.

Single QD [Fig. \ref{fig6} (b, d)]: The transmission zero of the
single QD is due to the destructive
interference of the two neighboring resonance states
\cite{lee,sadreev,mois} and is located between the energies of the
single QD. It is caused by the unitarity of the $S$ matrix with
account of the fact that the leads are attached to the
single QDs which are constituents of the  double dot system
\cite{rotter}. Around the energy $E_0$, the transmission amplitude
vanishes, $t(E_0) = 0$. Here Re$(t(E)) \sim (E - E_0)^2$ while
Im$(t(E)) \sim \frac{d ({\rm Re}(t(E))}{dE}\sim (E - E_0)$. It
holds therefore for the modulus of the transmission amplitude
$|t(E)| \sim |E-E_0|$ near $E_0$, see Fig. \ref{fig6}(b), and
$\frac{dt(E)}{dE}\mid_{E_0} \neq 0 $. Thus,  the phase of the
transmission amplitude arg$(t(E))$ jumps by $\pi$ at $E_0$
according  to \cite{mois}. The geometrical origin of this phase
jump can clearly be seen in Fig. \ref{fig6}(d). By pathing through
the origin of the coordinates Re$(t)=0$, Im$(t)=0$, the value
arg$(t)$  is not defined unambiguously. The phase jumps  by $\pm
\pi$, and the sign of the phase jump is not observable. We can
call such a zero  a first order zero.

Double QD [Fig. \ref{fig6} (a, c)]:
The zeros in the transmission through the double QD  with two
identical single QDs are of another type. They are of second order
since it is $|t(E)| \sim |t_1(E)|^2$ in the vicinity of the energy
$E = E_0$. It follows therefore  $t(E) \sim (E-E_0)^2$ near $E_0$,
see Fig. \ref{fig6} (a). The energy evolution of the real and
imaginary parts of the transmission amplitude is shown in Fig.
\ref{fig6}(c). In the inset of the figure, the evolution of
Re$(t),$ ~Im$(t)$ at the origin of the coordinate system is shown
in zoomed resolution. It has a cusp-like behavior, and there is no
phase jump at all. We present in Fig. \ref{fig6} (a) the phase
behavior of the transmission amplitude $t$ that is a combination
of two jumps with opposite sign, resulting in a zero phase jump at
the point $E=E_0$. This result agrees with the general statement
given by Barkay {\it et al} \cite{mois}, that phase jumps do not
appear when $\frac{dt(E)}{dE}\mid_{E_0} = 0 $ (as in our case) at
the energy $E = E_0$.

\begin{figure}[ht]
\includegraphics[width=.8\textwidth,height=0.2\textheight]{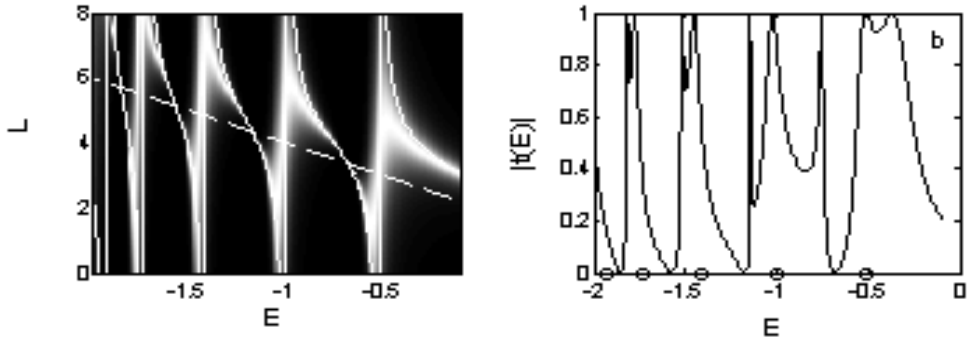}
\caption{(a) The  transmission through a double QD
 consisting of
two identical single QDs with $d=5$ states that are connected by a
wire, versus energy and length of the wire. The eigenvalues of
$H_B$ are shown by thin lines while the energy $
\epsilon(L)=1-L/2$   of the mode in the wire is shown by the
dashed line. ~$v=0.5, ~u= 0.2$. (b) Energy dependence of the
modules of the transmission amplitude  for $L = 4$. The energy
levels of the single QD are shown by circles at the energy axis,
$\varepsilon_L = -2\cos(\pi n/6), ~n = 1, 2,\ldots, 5$. }
\label{fig7}
\end{figure}

According to formula (\ref{T2}),  zeros of second order in the
single-channel transmission  of a double QD are given by  zeros of
first order in the transmission  of the single QDs (that
constitute the double QD) when they are identical. If  both single
QDs have $N$ energy levels then $N-1$ transmission zeros of second
order will appear in the double QD system. A numerical computation
for the particular case $N = 5$ confirms this conclusion (Fig.
\ref{fig7}).

\begin{figure}[ht]
\includegraphics[width=.8\textwidth,height=0.25\textheight]{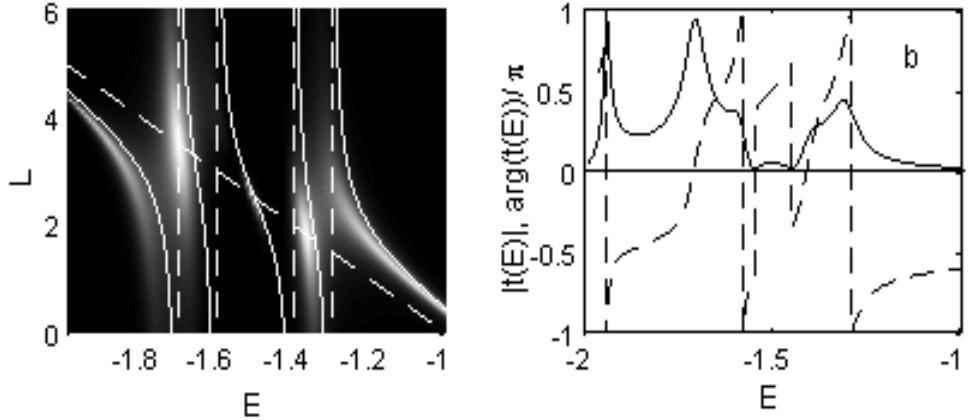}
\caption{(a) The  transmission   through a double QD
with different single QDs connected by a wire,
and (b) the energy dependence of the modules (solid line) and of
the phase $arg(t)/\pi$ (dashed line) of the
transmission amplitude for this double QD system at
a fixed length $L = 4$ of the wire. The
energy levels of the left single QD are $\varepsilon_k^L = -1.7,
-1.4$, while those  of the right single QD are $\varepsilon_k^R =
-1.6, -1.3.$. Further,  $\epsilon(L) = -1 -L/5, ~v=0.3, ~u = 0.1$.
The full lines in (a) are the eigenvalues of $H_B$ while the
dashed lines are the energies $\varepsilon_k^L, ~\varepsilon_k^R,$
and $\epsilon(E)$. } \label{fig8}
\end{figure}

Next we will consider the transmission through a double QD
consisting of two different single QDs coupled to the wire and to
the leads as shown in Fig. \ref{fig4}. When each single QD has $N$
states, the number of zeros in the transmission through the double
QD is, according to (\ref{T2}), ~$2(N-1)$. This conclusion is
demonstrated by the results of numerical calculations shown in
Fig. \ref{fig8}. Here, the two two-site single QDs are chosen to
be different from one another what can be achieved by either
different coupling constants $u$ between the single QDs and the
wire or by different energies of the levels of the two single QDs.
We have chosen the latter possibility.

In Fig. \ref{fig8}, we see two transmission zeros of first order.
The  phase jump is $-\pi$ at the first zero and  $\pi$ at
the second zero. Considering the transition $\varepsilon_k^l
\to\varepsilon_k^r$ with $k=1,2$, we see that the transmission zeros
will approach each other with the consequence that the  transmission
zero turns over into a second order zero. The phase jumps annihilate
each other as shown in Fig. \ref{fig6} (a).

\begin{figure}[ht]
\includegraphics[width=.6\textwidth,height=0.3\textheight]{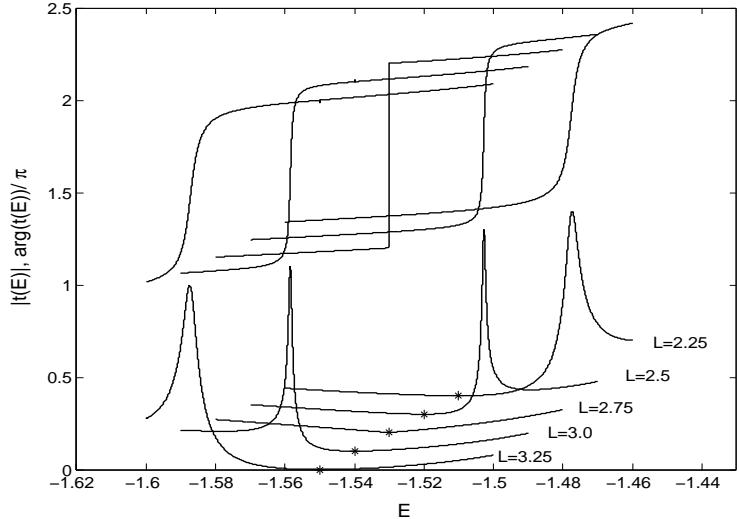}
\caption{ The energy dependence of the modules $|t(E)|$ (bottom)
and of the phase $arg(t(E))/\pi$ (top) of the
transmission amplitude for $L=$ 2.25, ~2.5, ~2.75, ~3.0, ~3.25.
The other parameters are the same as those in Fig. \ref{fig5}. The
transmission zeros are denoted by stars. They are of second order.
The ordinate is shifted every time by 0.1 when $ L$ is changed by
0.25. All phases are shifted by $\pi$. } \label{fig9}
\end{figure}

We consider now the evolution of the modules of the transmission
amplitude and the corresponding phase shifts when the decay width of
one of the states approaches zero. The results shown in Fig.
\ref{fig9}
are performed for the double QD system the transmission of which is shown
in Fig. \ref{fig5} together with the
eigenvalues $z_k$ of the effective Hamiltonian $H_{\rm eff}$, Eq.
(\ref{Heffgen}), as a function of $L$.
The latter ones are related to the poles  of the S-matrix. At $L =2.75$,
the third eigenstate crosses the energy of the transmission zero,
and its decay width Im$(z_3)/2$  approaches zero. As long as $L\ne 2.75$
and Im$(z_3) \ne 0$, the phase of the transmission amplitude
varies by $\pi$ more or less smoothly, according to the phase
shift caused by a resonance state with a  finite decay width. When
$L\rightarrow 2.75$ and Im$(z_3)\rightarrow 0$, the phase jumps by
$\pi$ due to the vanishing decay width of the resonance state. Therefore,
we have also in this case a phase jump of the transmission amplitude
by $\pi$. Correspondingly, the transmission zero
becomes of  first order at $L = 2.75$,  see  Fig. \ref{fig5}(b).
That means the resonance state whose decay width vanishes when
crossing the energy of the transmission zero,
restores the first order of the transmission zero  as well as the
phase jump by $\pi$.

The connection of the two single QDs to the leads and to the wire may
differ from that shown in Fig. \ref{fig4}. When one of the single QDs
loses its 2d-character by the manner it is integrated in the double
QD system, only  transmission zeros of first order appear in the
double QD. When both single QDs are included as 1d dots, the double
QD system has no transmission zero at all. In this case, the whole
system behaves as a 1d chain of sites without any transmission
zeros.

\section{Concluding remarks}

From the present study, we conclude that the simple  model used by
us for the description of a double QD system consisting of two
single QDs coupled by a wire, is efficient and describes
qualitatively the features of the resonant transmission through a
realistic double QD of the same structure. The difference between
the transmission through the realistic double QD shown in Fig.
\ref{fig1} and the model cases shown in Figs. \ref{fig3},
\ref{fig5}, \ref{fig7} and \ref{fig8} consists, above all, in the
fact that all features of the simple model are multiply repeating
in the realistic case. These multiple effects in  realistic cases
are related, obviously, to the large dimension of the single QDs
and to the higher number of eigenmodes  of the  wire inside
the double QD system.

The advantage of the simple model described by $S$ matrix theory
is that it allows a
clear discussion of the main features of the transmission,
especially of the transmission zeros and of the phase jumps
related to them. Of course, the model can give only qualitative
results that do not agree quantitatively  with the numerical
results (Fig. \ref{fig1}) obtained for the realistic double QD
system. However there is a large room to further develop the model.
First, we can take the eigenenergies of the wire for the single channel
propagation as
\begin{equation}\label{epsL}
  \epsilon_n(L)=\frac{\pi^2}{d^2}+\frac{\pi^2 n^2}{L^2}, n=1, 2,
  \ldots
\end{equation}
where $d$ is the width of the wire.
The first term in (\ref{epsL}) is related to the first-channel
propagation in the wire. Moreover we can take
into account that the eigenstates of the wire have the form
\begin{equation}\label{psiL}
  \psi_n(x)=\sqrt{\frac{2}{L}}\sin(\pi nx/L).
\end{equation}
Then $u$ can be calculated in the same way as $v$
\cite{dittes,sadreev}
\begin{equation}\label{uL}
  u_n(L)=u_0L^{-1/2}\psi_n'(x)\mid_{x=0,L} = u_n L^{-3/2},
\end{equation}
and the Hamiltonian (\ref{Hwire}) reads
\begin{equation}\label{Hw2}
H_w=\epsilon_1(L)|1\rangle \langle 1| + \epsilon_2(L)|2\rangle \langle 2|
\, .
\end{equation}

In our calculations we have chosen the widths of the wire inside
the double QD and the widths of leads attached to the QD to be the
same. Thereby, the leads and the wire both support only single
channel transmission through the double QD. Experimentally,
however, the width of the wire varies, usually, by applying  gate
voltages \cite{blick} while the length of the leads remains fixed.
In order to model this case, we can consider  two energy levels of
the wire one of which is related to the first channel propagation
(\ref{epsL}) and the other one  to the second channel propagation
\begin{equation}\label{epsL2}
  \epsilon_{n'}(L)=\frac{4\pi^2}{d^2}+\frac{\pi^2 n'^2}{L^2}, n'=1, 2,
  \ldots.
\end{equation}
When the wire's width $d$ exceeds the lead's width, we can take
one  energy level (\ref{epsL}) (say, $n=2$) together with another
one from (\ref{epsL2}) (say, $n'=1$) in such a manner that they
cross at a certain length $L_0$, ~$\epsilon_n(L_0)=
\epsilon_n'(L_0)$.

\begin{figure}[ht]
\includegraphics[width=.8\textwidth]{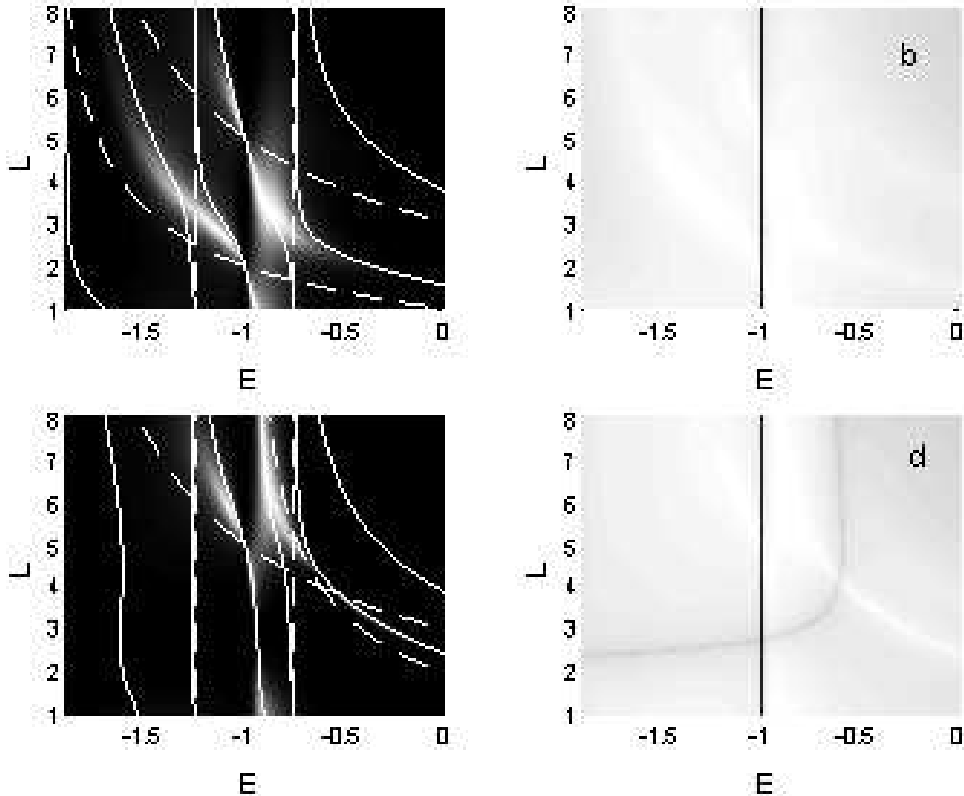}
\caption{
The  transmission probability (a,c) and its logarithm (b,d) for a
double QD with
a wire that has two energy levels. The single QDs have eigenenergies
$\varepsilon_k=-1.25, -0.75$. The coupling strengths $u$ between the
single QDs and the wire are taken according to (\ref{uL}), $u=0.2,
v=0.6$. In (a) and (b),
the first-channel propagation  is shown  with the two
energies of the wire given by  (\ref{epsL}),  $n=1, 2$. In (c) and (d),
the two channel propagation in the wire is shown with  the
energies $\epsilon_1(L)=4\pi^2/L^2$
and $\epsilon_2(L)=1+\pi^2/L^2$ that are crossing. The dashed lines
show the eigenenergies of the isolated wire and the single QDs while
the solid lines show the eigenenergies of the closed double QD system.}
\label{fig10}
\end{figure}

The numerical results obtained for the
transmission probability (Fig. \ref{fig10}) for such a case
are the following.
For   single channel propagation in the wire, the
transmission features including the transmission zeros are similar
to those discussed above.  The number of transmission zeros remains
unchanged, but the number of transmission peaks is
doubled corresponding to the two modes in the wire.
These results correspond to those obtained for
the single channel transmission through a realistic double dot system
(Fig. \ref{fig1}).
The situation is however another one when the wire can support two
channel propagation. In this case, the number of transmission
zeros is doubled. One of the lines of the transmission zeros does
not depend on the wire length $L$ (as in the model calculations of
Sect. IV), but the other one depends on $L$ and crosses the first
line of the transmission zeros. The multi channel
propagation induced by a variable width of the wire can lead,
therefore, to an essentially more complicated picture of the
transmission zeros. The picture of transmission zeros as a whole
remains however
comprehensible on the basis of the results obtained in the
framework of the simple model considered above.

It is experimentally  easier to vary the wire's width than its
length. It is important therefore to mention that the
width $d$ and length $L$ of the wire are equivalent for the
energy levels of the multi-channel wire according to Eqs.
(\ref{epsL}) and (\ref{epsL2}). Therefore, the transmission
probability versus energy and length shown in Fig. \ref{fig10} can
be qualitatively considered as the transmission probability versus
energy and width of the wire.

\acknowledgments  This work has been supported by RFBR grant
04-02-16408. A.F.S thanks also Max-Planck-Institut f\"ur Physik
komplexer Systeme for hospitality.\\

 $^{*}$ e-mails: rotter$@$mpipks-dresden.mpg.de;
almsa$@$ifm.liu.se, almas$@$tnp.krasn.ru,
almsa$@$mpipks-dresden.mpg.de

\end{document}